\documentclass[12pt]{article}
\usepackage[utf8x]{inputenc}
\usepackage{amsmath}
\usepackage{graphicx}
\usepackage{fullpage}

\title{Edge effects in game theoretic dynamics of spatially structured tumours}

\author{Artem Kaznatcheev \AND Jacob G. Scott \AND David Basanta}
\begin{document}

\begin{flushleft}

\Large

{\bf Edge effects in game theoretic dynamics of spatially structured tumours.}
\normalsize

\vspace{0.5cm}
Artem Kaznatcheev$^{1,2\dagger}$, Jacob G.\ Scott$^{2,3\dagger}$ \& David Basanta$^{2*}$ \\
\vspace{0.5cm}
$^1$ School of Computer Science and Department of Psychology, McGill University, Montreal, QC, Canada\\
$^2$ Integrated Mathematical Oncology, H. Lee Moffitt Cancer Center and Research Institute, Tampa, FL, USA\\
$^3$ Wolfson Centre for Mathematical Biology, Mathematical Institute, University of Oxford, Oxford, UK\\

\begin{abstract}
\textbf{Background:} 
Analysing tumour architecture for 
metastatic potential usually focuses on phenotypic differences due to cellular morphology or specific genetic mutations, but often ignore the cell's position within the heterogeneous substructure. 
Similar disregard for local neighborhood structure is common in mathematical models.

\textbf{Methods:} 
We view the dynamics of disease progression as an evolutionary game between cellular phenotypes.
A typical assumption in this modeling paradigm is that 
the probability of 
a given phenotypic strategy interacting with another depends exclusively on the abundance of those strategies without regard local heterogeneities. 
We address this limitation by using the Ohtsuki-Nowak transform to 
introduce spatial structure to the go vs. grow game. 

\textbf{Results:} 
We show that spatial structure can promote the invasive (go) strategy. 
By considering the change in neighbourhood size at a static boundary -- such as a blood-vessel, organ capsule, or basement membrane -- we show an edge effect 
that allows a tumour without invasive phenotypes in the bulk to have a polyclonal boundary 
with invasive cells. We present an example of this promotion of invasive (EMT positive) cells in a metastatic colony of prostate adenocarcinoma in bone marrow.

\textbf{Interpretation:} 
Pathologic analyses that do not distinguish between cells in the bulk and cells at a static edge of a tumour can 
underestimate the number of invasive cells.
We expect our approach to extend to other evolutionary game models where interaction neighborhoods change at fixed system boundaries.
\\
\end{abstract}
\textbf{keywords:} evolutionary game theory -- go-vs-grow game -- spatial structure -- heterogeneity -- edge effect -- EMT

\vfill
$^{\dagger}$ contributed equally\\
$^*$ artem.kaznatcheev@mail.mcgill.ca, jacob.g.scott@gmail.com, david@cancerEvo.org

\end{flushleft}




\newpage

\section{Introduction}

The importance of heterogeneity within tumours is gaining ground as one of the most important factors in the laboratory and clinic alike~\cite{hetero}. 
This heterogeneity exists at multiple scales, each of which presents its own unique set of challenges. 
One form of phenotypic heterogeneity that has been widely studied was first described by Giese et al.~\cite{Giese:2003ed} when they showed that in gliomas 
migration and proliferation were mutually exclusive processes; motile cells are unable to proliferate while they are moving and proliferating cells are unable to move while they divide. 
This has been termed the \textit{Go or Grow} dichotomy~\cite{Giese:1996nx}. 
A proliferative or autonomous growth (Grow) cell might switch to a motile or invasive (Go) cell either through mutation, metabolic stress~\cite{Godlewski:2010aa}, undergoing the Epithelial-Messenchymal Transition (EMT)~\cite{PW09}, or some other mechanism. 
EMT is generally characterised by the loss of cell-cell adhesion and a gain in in motility and invasiveness in tumour cells, and is one of the hallmarks of several carcinomas such as prostate~\cite{empP1,empP2}, breast~\cite{empBP}, and other ductal cancers where pre-invasive neoplasms are constrained by 
architectural boundaries (edges) such as the duct wall or basement membrane. 
Our goal in this work is to highlight the important and overlooked role of such edges in the evolutionary dynamics of the competition between Go and Grow cells.

Evolutionary Game Theory (EGT) is a mathematical approach to modeling frequency-dependent selection where players interact strategically not by choosing from a set of strategies but instead by using a fixed strategy determined by their phenotype. 
Given the evolutionary nature of cancer~\cite{Nowell:1976ul,Greaves:2012vn}, EGT has been applied to study how the interactions between different types of cells in a polyclonal tumour could drive the dynamics of a given cancer~\cite{Basanta:2008}. 
%
In its first application to oncology~\cite{Tomlinson:1997vm,Tomlinson:1997wn}, EGT was used to analyse the circumstances that lead to coexistence of two phenotypes. 
Subsequent research
~\cite{bach:2001fk} extended this idea to interactions between three players in the angiogenesis problem.
Gatenby and Vincent adopted a game theory model heavily influenced by population dynamics to investigate the influence of the tumour-host interface in colorectal carcinogenesis~\cite{gatenby:2003a,Gatenby:2005} and suggested therapeutic strategies~\cite{gatenby:2003b}. 
Our own work, as well as that of others, has shown that EGT can be used to study the conditions that select for more aggressive tumour phenotypes in gliomas~\cite{Basanta:2008gr,Basanta:2011hs}, colorectal cancer~\cite{gatenby:2003a,Gatenby:2005}, multiple myeloma~\cite{Dingli:2009en} and prostate cancer~\cite{Basanta:2011jb}. 
Furthermore, EGT has been used to investigate the impact of treatment on cancer progression~\cite{Basanta:2012ly,Orlando:2012zr,A13}. 
For an in depth overview of the game theoretical approach to cancer, see Basanta \& Deutsch ~\cite{Basanta:2008}.

\begin{figure*}[ht]
\centering
\includegraphics
[scale=0.386]{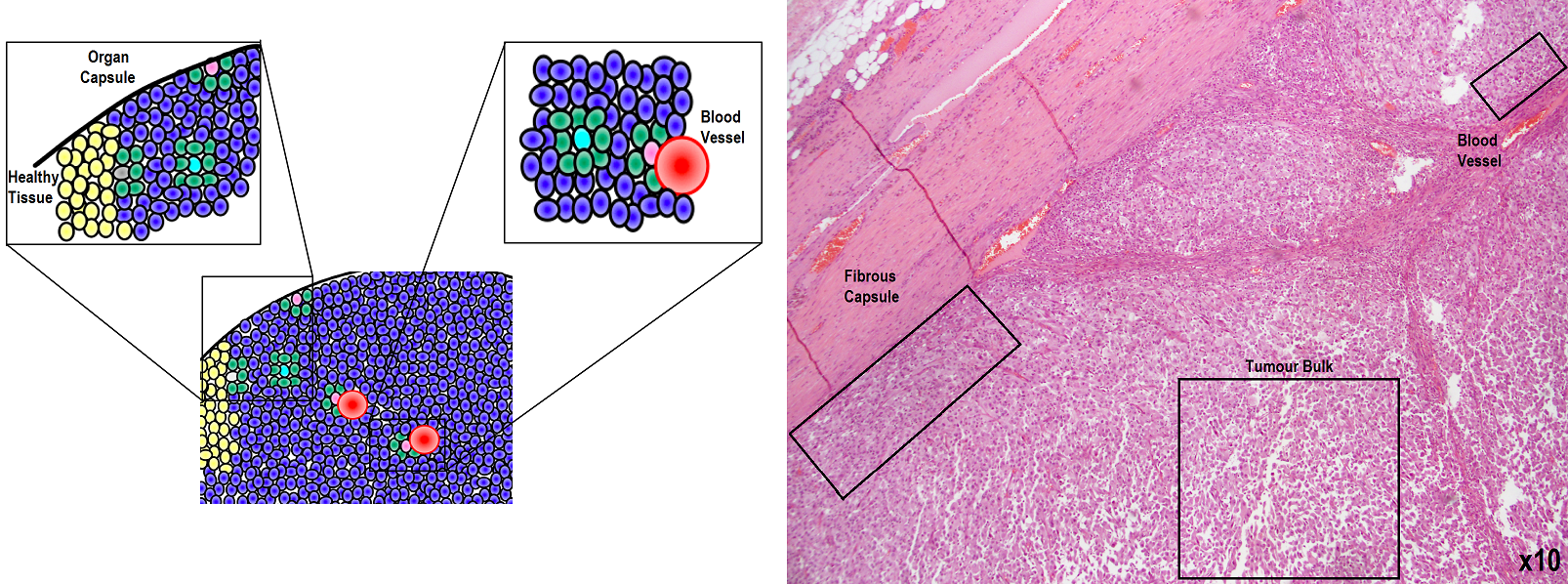}
\caption{
\textbf{An idealized image of a hypothetical tumour 
(Left),
and a clinically produced micrograph of a sarcoma under low power stained with Hematoxylin and Eosin 
(Right).
} 
A tumour cell (blue cells in cartoon) has several different scenarios that affect the architecture of its neighborhood geometry which we illustrate here. 
On short time scales, cells in a solid tumours experience largely static neighbourhood architectures, however, cells in the bulk of the tumour (turquoise in cartoon, and lower right box in the sample image) have many more neighbors than cells against static boundaries (pink) like an organ capsule (left close-up), fibrous capsule (left box in sample image), or blood vessel (right close-up in cartoon, upper right box in sample image).
This change in relative number of neighbours affects evolutionary game dynamics.
The boundary between the tumour and healthy cells (yellow), while of interest, is a dynamic edge and not considered in this article.
}
\label{fig:cartoon}
\end{figure*}

In this study, we introduce spatial structure into the canonical `go versus grow' game ~\cite{Basanta:2008gc,Hatzikirou:2010jk} in which proliferation and motility compete within a tumour.
We use our direct spatial approximation to consider a familiar scenario for conservation biology: the edge effect of an ecological system (for example, a forest in landscape ecology) at a static boundary~\cite{edgeEcologyDef,edgeEcologySurv,edgeEcologySci}.
In tumour progression, this is analogous to a cancer cell surrounded entirely by other cancer cells as opposed to constrained by a physical boundary, such as a basement membrane, organ capsule, fibrous capsule, or blood vessel (see Figure~\ref{fig:cartoon}).
Note that we do not consider the unconstrained, dynamically growing edge of the tumour.

The static boundaries studied in this article are exciting in and of themselves, since the evolutionary dynamics that occur at static boundaries govern progression past key cancer stages: the change from \textit{in situ} to invasive; locally contained to regional advanced growth; and the dramatic shift from local to metastatic disease.
The former situation occurs early in the progression of most epithelial tumours, and it is commonly believed that it is at this point when the Warburg effect occurs, pushing cells toward the glycolytic (acid producing) phenotype which promotes invasion and motility - the so called `acid mediated invasion' hypothesis~\cite{Gatenby:2004fc,Gatenby:2006je,Gillies:2008wm,Basanta:2008gr}. 
The latter situation occurs when tumours are up against blood vessels and is likely the first opportunity for hematogenous dissemination, the first step in the metastatic process~\cite{Chambers:2002tv}.
Here, we show a striking change in the evolutionary game dynamics from the tumour bulk to the tumour's physical boundary (Figure~\ref{fig:cartoon}).  
This study  represents, to our knowledge, the first attempt to understand the effects of changing neighborhood structure on evolutionary game dynamics of tumours.

\section{Methods}


\subsection{Inviscid game for motility}
\label{sec:inviscid}

To mathematically model the \textit{Go or Grow} dichotomy~\cite{Giese:1996nx} we considered the situation in which the tumour is made of a population of rapidly proliferating cells capable of autonomous growth (AG), along with a subpopulation which arises by a mutation or phenotypic change which confers motility/invasiveness (INV) to tumour cells. 
The game has two parameters: $c$ represents the direct and indirect costs of motility incurred by  cells with the INV phenotype resulting from the reduced proliferation rate of 
motile/invasive cells~\cite{Giese:1996nx} and $b$ is the maximum fitness a tumour cell will have under ideal circumstances when it does not have to share space or nutrients with other cells. 
As the units of measure are arbitrary, the ratio $c/b$ can be considered alone to determine game theoretic dynamics.  
With INV as strategy 1 and AG as strategy 2, the game's payoff matrix is:

\begin{equation}
\bordermatrix{ ~ & \text{INV} & \text{AG} \cr
\text{INV} & \frac{1}{2}b + \frac{1}{2}(b - c) & b - c \cr
\text{AG} &	b & \frac{1}{2}b \cr
}.
\label{eq:game}
\end{equation}

\noindent To understand the inviscid model~\cite{Basanta:2008gc}, imagine two cells meeting at random in a resource spot; 
for an inviscid population this probability depends only on the cells' relative abundance but for the structured populations considered later, the probabily of meeting again will be higher. 
If both cells are INV (motile) then one cell stays in the resource spot (i.e. a  location that contains oxygen, glucose or other growth factors) and consumes the resources $b$, and the other pays a cost, $c$, to move and find a new empty site where it can then find resource, for a payoff $b$. 
If after a move, a motile cell only finds a new empty spot with probability $r$, then the expected move payoff is $rb - c$. 
This can be captured as an indirect cost by adjusting the cost to be $c' = c + (1 - r)b$ without introducing an extra parameter. 
As the cell that has to move is chosen randomly from the two, the expected payoff for each cell is the average of the no-move ($b$) and move ($b - c$) payoffs. 
On the other hand, if an INV cell meets an AG cell then the INV cell will move, incurring the cost, $c$, before receiving the benefit, $b$, for a total fitness ($b - c$). 
The AG cell, however, will stay and consume all the resources ($b$).
Finally, if two non-motile cells (AG) are in the same resource spot then they simply share the resources, for a payoff of $b/2$. 

\subsection{Direct approximation of spatial structure}

A standard assumption in EGT is a perfectly mixed (inviscid) population, in which every cell in the population is equally likely to interact with every other cell~\cite{mathOverview}. 
This may be a reasonable assumption in liquid tumours, but in solid tumours (or any other situation being modeled) in which spatial structure is important, the validity of this assumption should be questioned~\cite{SF07}. 
The current solution to this is to map analytic EGT cancer models onto a lattice and run \emph{in-silico} experiments to simulate the resulting Cellular Automaton~\cite{bach:2003,mansury:2006,Basanta:2008gc} 
In such cases, the choices of the specific microdynamics to simulate are arbitrary and often left up to convention and the modeler's imagination, since direct empirical mechanisms at such a precise level are often unknown.
Further, explicitly solving complex spatial structures is currently outside of existing mathematical tools, so computational approaches have to sacrifice the analytical power and full theoretical understanding of pure EGT approaches.

On occasion, computational modelers restore some analytical power by making mathematical approximations of the simulation that are already approximations of, or guesses at, the tumour's real spatial structure. 
To avoid this double approximation and to analytically model how spatial structure effects evolutionary games in the limit of large populations and weak selection, Ohtsuki \& Nowak \cite{Ohtsuki:2006vn} derived a simple rule for taking a more direct first-order approximation of any spatial structure. 
This approach is based on the technique of pair-approximation~\cite{MTSO87,MOSS92,vB00} and is exact only for Bethe-lattices (infinite trees of constant degree), but is highly accurate for any static structure where higher-order terms, like the correlations between neighbours of neighbours, are negligible. 
For example, Ohtsuki and Nowak concentrated on the application of their tool to $k$-regular random graphs, which are locally tree-like and have negligible second-order and higher terms, but the transform can be used more broadly. 
While real spatially structured biological populations, such as solid tumours, can have non-negligible higher-order interactions, Ohtsuki and Nowak's first-order approximation is an improvement over the common inviscid assumption that still allows us to explore a completely analytic model.

Given a game matrix $A$, one can compute the Ohtsuki-Nowak (ON) transform $A' = \text{ON}_k(A)$ and then recover the dynamics of the spatially structured game $A$ by simply looking at the inviscid replicator equation of $A'$. Here, we present the transform in a form that stresses its important qualitative aspects:


\begin{equation}
	ON_k(A) = A + 
    \underbrace{\frac{1}{k-2}(\vec{\Delta}\vec{1}^T - \vec{1}\vec{\Delta}^T)}_\text{local dispersal} \; + \; \underbrace{\frac{1}{(k + 1)(k - 2)}(A - A^T)}_{\substack{\text{finite sampling from} \\ \text{death-birth updating}}},
    \label{eq:ONtrans}
\end{equation}

\noindent where $\vec{1}$ is the all ones vector and  $\vec{\Delta}$ is the diagonal of the game matrix $A = [a_{ij}]$, i.e. $\vec{\Delta}_i = a_{ii}$; 
thus,  $\vec{\Delta}\vec{1}^T$ ($\vec{1}\vec{\Delta}^T$) is a matrix with diagonal elements repeated to fill each row (column). 
The first summand is the original payoff matrix.
The second summand accounts for the more common same-strain interactions that are a consequence of local dispersal.
The type of perturbation in the third summand was shown by  Hilbe~\cite{Hilbe:2011dq} to result from finite sampling of interaction partners.
The summands are not arbitrary, and emerge as a whole from the pair-approximation technique. Our rationale forgrouping the summands in this particular form is to help build intuition for equation~\ref{eq:ONtrans}. 
Effects of indirect interaction from more distant cells could be introduced as further corrections to Equation~\ref{eq:ONtrans}, but would require more assumptions about the tumour microstructure~\cite{vB00}.
Since the local neighbour relation is empirically well studied~\cite{calvo2011cell}, the focus of the current study is incorporating only this first-order structure into an EGT model of cancer dynamics.

\section{Results}

Whenever $b > c$, the game in equation~\ref{eq:game} is a social dilemma (like the Prisoner's dilemma or Hawk-Dove game, for a classification see~\cite{S10})  with invasive cells as the cooperators, and AG cells as defectors (mixing invasive and AG).
Rules of thumb from EGT~\cite{Ohtsuki:2006kx} suggest that cooperators benefit from the structure of small interaction neighbourhoods, in agreement with our biological intuition that, in this game, having the ability for conditional motion is of more use in a more constrained and viscous environment than in one where all cells are already stochastically moving around and interacting at random. 
We look at this formally by explicitly considering spatial effects on the previously inviscid model. 
Applying the transform from equation~\ref{eq:ONtrans} to the game in  equation~\ref{eq:game} yields:

\begin{equation}
\begin{pmatrix}
\frac{\displaystyle 1}{\displaystyle 2}b + \frac{\displaystyle 1}{\displaystyle 2}(b - c) 
& b\frac{\displaystyle 2k - 3}{\displaystyle 2(k - 2)} - c\frac{\displaystyle 2k^2 - k - 1}{\displaystyle 2(k - 2)(k + 1)} \\
b\frac{\displaystyle 2k - 5}{\displaystyle 2(k - 2)} + c\frac{\displaystyle k + 3}{\displaystyle 2(k - 2)(k + 1)} 
& \frac{\displaystyle 1}{\displaystyle 2}b
\end{pmatrix}.
\label{eq:ONgame}
\end{equation}

\noindent This game has three qualitatively different regimes that depend on the value of $\frac{c}{b}$ and $k$:
\begin{enumerate}
\item If $\frac{k + 1}{k^2 + 1} \geq \frac{c}{b}$ then there is a single stable fixed point with all cells invasive. 
All polyclonal tumors evolve toward this fixed point. 
For inviscid populations ($k \rightarrow \infty$) this condition is satisfied only if motility is cost-free ($c = 0$) and hence the possibility of an all invasive tumour was not noticed in previous non-spatial analysis~\cite{Basanta:2008gc}.
\item If $\frac{k + 1}{k^2 + 1} < \frac{c}{b} < \frac{k + 1}{2k + 1}$ then the game has Hawk-Dove dynamics and there is a stable polyclonal equilibrium with a proportion $p$ of INV cells:
\begin{equation}
p = \frac{b - 2c}{b - c} + \frac{1}{k - 2} - \frac{1}{(k + 1)(k - 2)}\frac{2c}{(b - c)}.
\end{equation}
All polyclonal populations will converge toward this proportion of INV cells. 
In the unstructured limit as $k \rightarrow \infty$ we have perfect agreement with our previous results~\cite{Basanta:2008gc} and recover the condition $\frac{c}{b} \leq \frac{1}{2}$ that was assumed for the inviscid  equilibrium to exist and the exact numeric value of $\frac{b - 2c}{b - c}$ for the equilibrium proportion of INV agents.
For any finite $k$, however, the proportion of invasive cells is strictly higher.
\item If $\frac{k + 1}{2k + 1} \leq \frac{c}{b}$ then the game has Prisoner's dilemma dynamics and any polyclonal population converges toward all AG and the tumor remains non-invasive.
\end{enumerate}

\begin{figure*}
\centering
\includegraphics[scale=0.723]{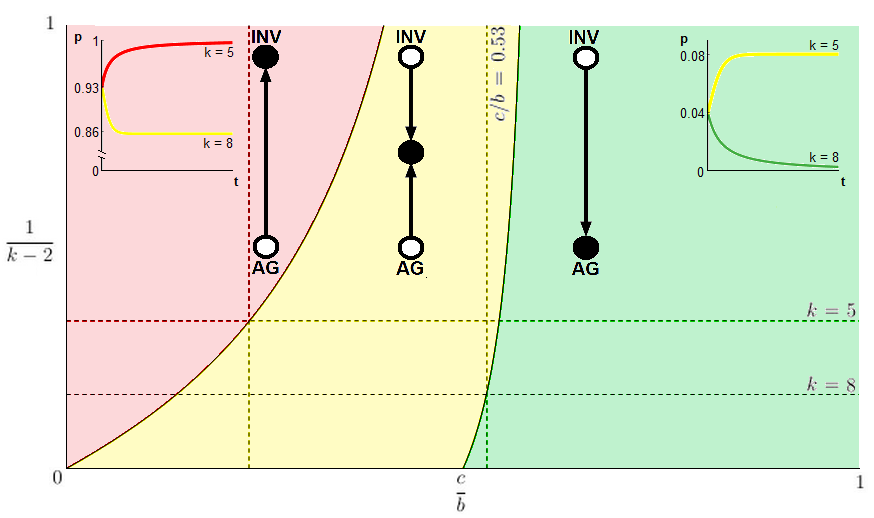}
\caption{
\textbf{Evolutionary game dynamics as a function of changing neighborhood size and relative cost of motility.} Here, we plot level of viscosity $\frac{1}{k - 2}$ versus relative cost of motility $\frac{c}{b}$. 
The parameter space is divided into three regions that correspond to qualitatively different dynamics. 
In the red, the population evolves toward all INV; 
in the yellow -- toward a polyclonal tumor of INV and AG cells; 
and in the green the tumor remains all AG. 
When $\frac{1}{k - 2} = 0$ (i.e. $k \rightarrow \infty$) we recover the standard inviscid replicator dynamics of our previous work~\cite{Basanta:2008gc}. 
For $\frac{1}{k - 2} = 1$ ($k = 3$, the top edge of the plot), we have the environment with the smallest local neighbourhood to which the ON-transform applies. 
The first horizontal dotted line marks $k = 5$ (pink cell in fig.~\ref{fig:cartoon}) and the bottom line is $k = 8$ (teal cell in fig.~\ref{fig:cartoon}). 
The left vertical dotted line is at $c/b = 0.23$, and shows that it is possible to go from a polyclonal tumor in the bulk to a completely invasive population at a static edge.
The right vertical dotted lines shows that is possible to see a qualitative shift from all AG to a polyclonal tumour in dynamics with the game fixed at $c/b = 0.53$ by decreasing $k$ from $8$ and $5$ (increasing $\frac{1}{k - 2}$ from $1/6$ to $1/3$) at the tumour boundary.
Example dynamics from a numerical solution of the replicator equation of the transformed game are shown in the insets, where proportion of INV (p) is plotted versus time (t).
The equation specifying the dynamics is $\dot{p} = p((ON_k(G)\vec{p})_{1} + \vec{p}^T ON_k(G) \vec{p})$ where $\vec{p}^T = ( p \;,\;\; 1 - p )$, $G$ is the game in eq.~\ref{eq:game}, and $ON_k$ is the transform from equation ~\ref{eq:ONtrans}.
The left inset corresponds to $c/b = 0.23$, an initial proportion of invasive agents of $p_0 = 0.93$, $k = 5$ (tumour edge) for the red line, and $k = 8$ (tumour bulk) for the yellow.
The right inset corresponds to $c/b = 0.53$, an initial proportion of invasive agents of $p_0 = 0.04$, $k = 5$ (tumour edge) for the yellow line, and $k = 8$ (tumor bulk) for the green.}
\label{fig:params}
\end{figure*}

\noindent These three regimes are plotted in Figure~\ref{fig:params}. 
When $\frac{1}{k - 2} = 0$, we have the inviscid game and for $\frac{1}{k - 2} = 1$ we have the most structured regime possible with small neighbourhoods ($k = 3$). 
The red region corresponds to a completely INV tumour, the yellow to a polyclonal tumour, and the green to all AG. 
As we make the tumor more structured and reduce the number of neighbours, it becomes easier for the INV cells to be expressed in the tumor. 

\begin{figure*}[ht]
\centering
\includegraphics[scale=0.478]{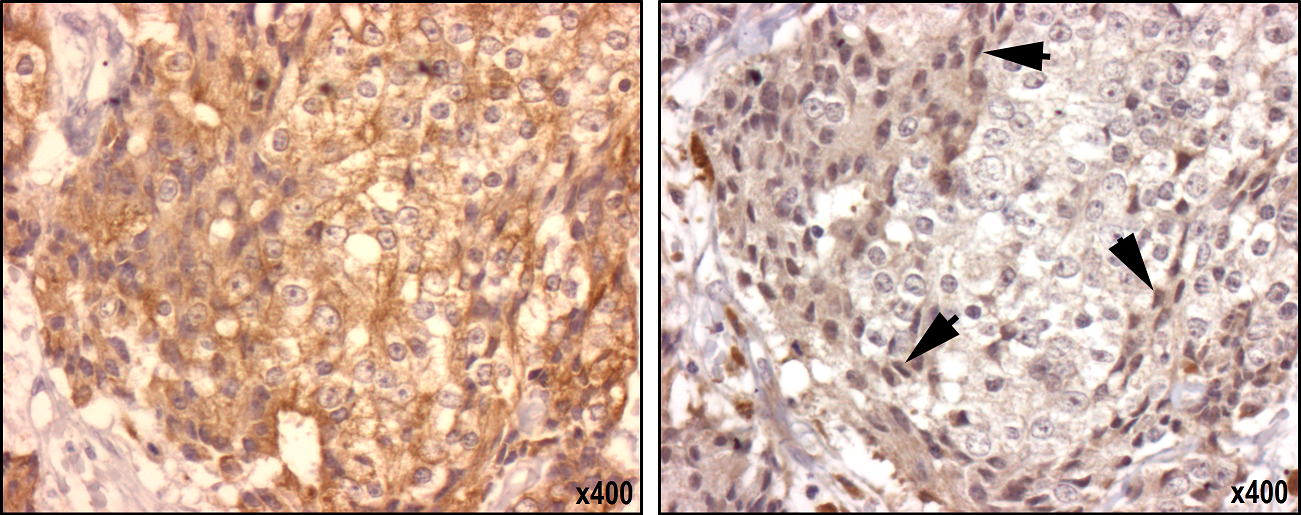}
\caption{
\textbf{EMT is found to be upregulated near static boundaries within bone marrow in bony metastatic deposits of prostate cancer.} 
In serial sections taken of bony metastatic deposits of prostatic adenocarcinoma at rapid autopsy, staining for SLUG is found to be increase near to the static boundaries created by the bony trabeculae.
Nearly homogeneous pancytokeratin staining (Left) reveals the ubiquity of metastatic carcinoma in this sample.
In the next slice, stained for SLUG (a marker of EMT), we see increased uptake of the antibody in the carcinoma cells lining the static boundary, suggesting a more motile phenotype (highlighted by arrows) as compared to those in the neighborhood representing the tumour bulk.
}

\label{fig:SLUG}
\end{figure*}

For example, consider the case where $\frac{c}{b} = 0.53$, and as in Figure~\ref{fig:cartoon}, in the tumor away from boundary (the teal cell) there are $k = 8$ neighbours and the dynamics favour all AG, so no INV cells will be present at equilibrium.
For example replicator dynamics of this condition, see the green line of the right inset.
If the solid tumor is pressed up against a static boundary then cells at the edge have fewer neighbours (e.g. $k = 5$, the purple cell) and the dynamics at the boundary favour a polyclonal population with about 8\% of the cells INV.
For example replicator dynamics of this condition, see the the yellow line of the right inset. 
Notice that the tumours represented by the green ($k = 8$) and yellow ($k = 5$) lines of the right inset have the same $c/b = 0.53$ and initial proportion of invasive cells $p_0 = 0.04$ and yet the invasive phenotype is pushed to extinction in the tumour bulk ($k = 8$) but stabilizes near potentially dangerous level of invasive cells ($p = 0.08$) at the tumour edge ($k = 5$). 
Similar higher selection for invasiveness at the static edge is present for the more competitive environment of $\frac{c}{b} = 0.23$ in the left inset, but in this case we started the example replicator dynamics at $p_0 = 0.93$. In this case, we have a polyclonal tumour bulk ($k = 8$) with $p = 0.86$ and convergence towards all invasive phenotypes at the static edge ($k = 5$).
Of course, the specific parameter values above are for illustrative purposes.
Actual change in $k$ (just like $c/b$) will be experiment and geometry dependent -- for example, we expect a more drastic decrease in $k$ for a convex rather than concave boundary. 
The main message is that the edge effect can cause a polyclonal boundary in a tumor with a homogeneous all AG body.
And while not yet shown to be universally applicable, this is in qualitative agreement with the experience of pathologists, such as the typical staining for motility (EMT) in Figure~\ref{fig:SLUG} where we see the bone marrow space entirely invaded by carcinoma cells (Pancytokeratin, Left), but a stark difference in motile phenotype as illustrated by the nuclear staining by SLUG, which is only present in high concentration in cells against the static boundary of the bony trabecula.


\section {Discussion}
A standard assumption in evolutionary game theory is that all players interact with all other players: the population is inviscid. 
There are a number of biological scenarios in which such an assumption could be misleading, and we consider such a scenario in the form several key aspects of solid tumour progression. 
The role of spatial heterogeneity has been explored before in mathematical models studying evolutionary processes in cancer~\cite{Anderson:2006wo,Komarova:2013jk,Thalhauser:2010qy}. 
Those models show that different environments produce different selective pressures and that phenotypic heterogeneity results from, and drives, the spatial one. 
Our approach tackles a different question related to the nature of physical edges on cancer evolution for which we have applied the Ohtsuki-Nowak transform~\cite{Ohtsuki:2006vn} to the standard go-grow game of mathematical oncology~\cite{Basanta:2008gc}.
We have shown a quantitative effect:  spatial-structured tumours promote the invasive phenotype compared to inviscid tumours with  the same ratio of cost of motility to benefit of resources $c/b$.

We also considered the decrease in neighbourhood size experienced by cells at the static boundary of a tumour, compared to cells within the tumour bulk. 
This could represent a number of feasible and very relevant scenarios, including a tumour against a blood vessel, organ capsule, or fibrous capsule; or an \textit{in situ} neoplasm at the basement membrane.  
We have shown that this change in neighbourhood for tumour cells can, independently from any other parameters, significantly effect the evolutionary stable strategies: in this case a promotion of the INV phenotype. 
The edge effect allows a tumour that internally has no invasive phenotypes expressed to have a polyclonal boundary with both invasive and non-invasive cells, a scenario which is known to appear in several cancers, most notably prostate \cite{villers1989role} in the form of peri-neural invasion, and the recently described perivascular invasion in melanoma~\cite{bald2014ultraviolet} (previously called angiotropism). 
In each of these two scenarios tumour cells express the motile phenotype, but only when against the physical structure in question.  
Another recent result suggesting the importance of effects of changes in local architecture is the work of Belmonte-Beitia et al.~\cite{belmonte2013modelling} in which the authors show that the dynamics of motility change drastically at the grey-white interface in glioblastoma in ways that can not be predicted by simply comparing the beheaviour in each zone individually.

The results of our mathematical model could have significant translational implications. 
Genetic heterogeneity has recently become recognized as the rule in cancer~\cite{Gerlinger:2012fs}, but as long as physicians have had microscopes, we have realised that spatial organisational heterogeneity was an  equally defining factor. 
The Gleason score~\cite{Gleason:1992kx} is a classic example of greater heterogeneity predicting worse survival in prostate cancer.
We have shown that a specific change in neighbourhood size at a static boundary can dramatically alter game dynamics, and select for novel phenotypes, and gives a rationale for working to understand within-tumour differences, not just at relatively long length scales~\cite{sottoriva2013intratumor}, but also at architecturally different locations at short length scales, which can be done with the advent of single cell technologies and laser capture microdissection.

We have shown that the local spatial structure of a tumour can strongly affect the evolutionary pressures on its constituent cells, even if all other factors are held constant.  
This can add yet another source of sampling bias to tissue biopsies and suggests that the architectural, not just the molecular, context is important.
For instance, consider an idealized fine-needle aspiration biopsy~\cite{biopsyOld}, assuming the standard 0.7 mm needle samples a perfect column around 20 cells in diameter of tumour cells right next to a critical boundary such as a capillary.
In our running example of $c/b = 0.53$ (see figures~\ref{fig:cartoon} and \ref{fig:params}) this would result in the sample containing only about $0.4\%$ invasive cells since 19 out of every 20 cells are not at the boundary.
This is below the detection levels of the state of the art medical practice~\cite{biopsy2009,biopsy2010}.
However, the critical 1 out of every 20 cells at the boundary, would have a dangerous $8\%$ invasive cells. 
Thus, an oncologist performing a diagnostic fine-needle aspiration biopsy could be led to think that a tumour poses a low risk for invasion/metastasis because the technique destroys the local structure of the tumour and mixes the cells at the critical tumour boundary with the (in this case) irrelevant tumour bulk.

Our model was motivated by the study of cancer, but the spatial edge effects in games of interacting players that we investigate could represent any number of other scenarios. 
While we have focused on the specifics of metastasis and cancer invasion, this method could yield insights into many other interesting problems ranging from ecology to medicine, and highlights the importance of neighborhood  geometry when studying the evolutionary dynamics of competing biological agents.

\section*{Acknowledgements} 
We would like to thank Dr. Colm Morrissey from the University of Washington and Dr. Marilyn Bui from the Moffitt Cancer Center for illustrative images used in this paper. 
DB would like to acknowledge U01CA151924-01A1 for financial support. JGS would also like to thank the NIH for its generosity in providing a Loan Repayment Grant.
\bibliography{gogrow}
\bibliographystyle{unsrt}

\end{document}